\documentclass[11pt]{article}
\usepackage{graphicx}

\newsavebox{\hflrar}
\sbox{\hflrar}{\makebox[0pt][l]
{${\scriptstyle \leftharpoonup}$}{${\scriptstyle \rightharpoonup}$}}

\def \to {\rightarrow}

\begin{document}
\pagestyle{plain}
\vskip 10mm
\begin{center}
{\bf\Large Predictions for $e^+e^-\to J/\psi \eta_c$  with Light-Cone Wave-Functions} \\
\vskip 10mm
J.P. Ma   \\
{\small {\it Institute of Theoretical Physics , Academia
Sinica, Beijing 100080, China \ \ \ }} \\
{\small {\it Department of Physics, Shandong University, Jinan Shandong 250100, China}}
\\
Z.G. Si \\
{\small {\it Department of Physics, Shandong University, Jinan Shandong 250100, China}}
\end{center}
\vskip 0.4 cm

\begin{abstract}
Predictions for $e^+e^-\to J/\psi \eta_c$ from previous studies are made by taking charmonia as
a nonrelativistic bound state and by using nonrelativistic QCD(NRQCD) approach. The predicted
cross-section is smaller by an order of magnitude than the experimentally observed.
We study the process by taking charm quark as a light quark and use light-cone wave-functions
to parameterize nonperturbative effects related to charmonia. The total cross section
of $e^+e^-\to J/\psi \eta_c$ can be predicted, if these wave-functions are known.
Motivated by studies of light-cone wave-functions of light hadrons, we make
a reasonable assumption of the forms of light-cone wave-functions.
With these light-cone wave-functions
we can obtain the cross section
which is more closer to the experimentally observed than that from NRQCD approach.
We also discuss in detail the difference between two approaches.
\end{abstract}
\par\vfil\eject

\par\vskip20pt
At Belle the production of double charmonia at $e^+e^-$ collider with $\sqrt{s}=10.6$GeV
has been studied. The experimental
result is given as\cite{Belle}:
\begin{equation}
\sigma(e^+e^-\to J/\psi\eta_c){\rm Br}(\eta_c\to {\rm 4\ charged\ particles}) = 33^{+7}_{-6}\pm 9 {\rm fb}.
\end{equation}
Since a branching ratio is smaller than 1, the above experimental result gives a lower
bound for the cross section. This experimental result is in conflict with theoretical predictions.
Theoretical predictions are made by taking a charmonium as a bound state of a $c\bar c$ quark.
Employing nonrelativistic wave-functions for such a bound state one can predict
production rates like the one measured at $e^+e^-$ colliders. Starting from this, the process
were studied in \cite{Ktchao,BraLe,Qiao}. From these studies
the cross-section is  about  $2\sim 5$fb. Comparing with Eq.(1) the experimentally measured cross section is
about an order of magnitude larger than theoretical predictions.
\par
If one takes charm quarks as heavy quarks,  a charmonium system
can be thought as a bound state consisting of a $c\bar c$ quark
mainly, in which the $c$- and $\bar c$-quark moves with a small
velocity. This fact enables us to describe such a system by an
expansion in the small velocity. A systematic expansion can be
achieved by using nonrelativistic QCD(NRQCD)\cite{NRQCD}. Within
this framework inclusive decays and inclusive productions of
single quarkonium can be studied consistently and rigorously,
where a factorization of nonperturbative effects can be completed.
But it is expected that theoretical predictions for charmonia can
have large uncertainties in general.
There are two important sources of corrections. One is
of relativistic correction.
Because the velocity of a $c$
quark in a charmonium is not very small, the relativistic correction
is large. This has been shown in different processes studied in
\cite{Gremm:1997dq,Bodwin:2003wh,Ma:2002ev}. Another source of correction is from
high order of $\alpha_s$. If one works with NRQCD factorization
for the process $e^+e^-\to\eta_c J/\psi$, the charm quark mass $m_c$ can not
be neglected and large logarithms like $\ln(m_c^2/s)$ will appear.
Those large logarithms can spoil the perturbative expansion in $\alpha_s$
and a resummation is needed.
However it is not expect that these large
uncertainties can result in such a large discrepancy. To explain
the discrepancy it was suggested that the experimental signals for
the final state of $J/\psi\eta_c$ may contain those of double
$J/\psi$\cite{BoBL} and the initial state interaction can enhance
the cross section of $J/\psi\eta_c$\cite{Luch}. Indeed, the
cross-section becomes large if these suggested effects are taken
into account. But, the discrepancy still remains large.
\par
It should be noted that for a process involving two quarkonia there is no rigorous theory based
on NRQCD in the sense of factorization of nonperturbative effects. Although a proof of the
factorization is not given yet, but it can be the case that the factorization does not hold.
The reason is the following: NRQCD is only applicable for a quarkonium in its rest frame.
For a process involving only one quarkonium one can always find such a rest frame through
a Lorentz boost. While for a process involving two or more quarkonia one can not
find in general a frame in which all quarkonia are in rest.
Therefore, a complete factorization may not be achieved.
\par
If the center-mass energy $\sqrt{s}$ is very large, i.e., $\sqrt{s}\gg m_c$,
one can take $c$-quark as a light quark. Then one can use light-cone wave-functions
to describe
nonperturbative effects of charmonia and a factorized form of the production amplitude
in terms of these wave-functions and a perturbative part
can be obtained. Such an approach for exclusive processes was proposed long time ago\cite{BL}.
Recently, light-cone wave-functions have been employed for charmonia to study their production
in B-decay\cite{cheng, song} and in photoproduction\cite{photoj}.
In comparison with the approach based on NRQCD for the process
$e^+e^-\to J/\psi\eta_c$, where the expansion parameter is the velocity, the approach with
light-cone wave-function is with the expansion parameters as $\Lambda/\sqrt{s}$, where $\Lambda$
is a soft scale and can be $\Lambda_{QCD}$, $m_c$ and masses of charmonia.
In this letter we will use
this approach to study the process $e^+e^-\to J/\psi\eta_c$. We will work at
the leading order of the expansion.
The correction to our result is at order of $\Lambda/\sqrt{s}$ or
$(\Lambda/\sqrt{s})^2$. Taking $\Lambda$ to be the mass of $J/\psi$, one can expect
that our result at $\sqrt{s}=10$ GeV has an uncertainty at level of $30\%$ or smaller.
Since an expansion in masses is used, there is only one large scale $\sqrt{s}$
in the perturbative part
and our prediction will not contain large logarithms like $\ln(m_c^2/s)$ if one takes
higher orders in $\alpha_s$ into account. However, such large logarithms
like $\ln(m_c^2/s)$ and $\ln(\Lambda^2_{QCD}/s)$
will appear in light-cone wave-functions. These large logarithms can be resummed
with evolution equations of wave-functions, like the Efremov-Radyushkin-Brodsky-Lepage
evolution equation\cite{ERBL}.
\par
In principle the $c\bar c g$ components of charmonia will also contribute, if one
take light-cone wave-functions at twist-3 into account.
Although NRQCD factorization may not directly
be applicable to the process studied here, but it can be used
to study light-cone wave-functions of charmonia as those introduced below,
in which nonperturbative effects can be factorized into NRQCD matrix elements
and light-cone wave-functions will be proportional to NRQCD matrix elements\cite{MaF}.
The light-cone wave-functions of $c\bar c g$ components will be proportional
to NRQCD matrix elements containing gluon fields explicitly.
It is known
that NRQCD matrix elements containing gluon fields is small by
NRQCD power counting\cite{NRQCD}, although they can be
important in some processes because of some mechanism of enhancement.
In our approach such an enhancement does not exist, we hence neglect contributions
from these $c\bar c g$ components.
We will only consider contributions from $c\bar c$ components.
\par
We consider the exclusive process:
\begin{equation}
 e^+ (p_1) + e^-(p_2) \to \gamma^*(q) \to J/\psi (p) + \eta_c(k),
\end{equation}
where momenta are indicated in the brackets. The amplitude can be
written as:
\begin{equation}
{\mathcal T} = i
           e \bar u (p_1) \gamma_\mu v(p_2) \frac{1}{q^2} \varepsilon^{\mu\nu\alpha\beta}
                   \varepsilon^*_\nu (p) p_\alpha k_\beta {\mathcal F}(q^2),
\end{equation}
where $\varepsilon^*_\nu (p)$ is the polarization vector of
$J/\psi$ and ${\mathcal F}(q^2)$ is the form factor defined as
\begin{equation}
  \langle J/\psi (p) \eta_c(k) \vert J^\mu \vert 0\rangle =i Q_c e \varepsilon^{\mu\nu\alpha\beta}
                   \varepsilon^*_\nu (p) p_\alpha k_\beta {\mathcal F}(q^2),
\end{equation}
where $Q_c$ is the charge fraction of $c$-quark in unit of $e$. From the
above the produced
$J/\psi$ is transversally polarized. With the form factor the
cross section can be calculated as:
\begin{equation}
\sigma(e^+e^-\to J/\psi \eta_c) =4\pi \alpha^2 Q_c^2 \frac{ \vert {\mathcal
F}(s)\vert^2 } {64} \left (1- \frac{4m_h^2}{s}\right
)^{\frac{3}{2}} \int_{-1}^{+1} d x (1+x^2),
\end{equation}
where $x=\cos\theta$ and $\theta$ is the angle between $J/\psi$
and $e^+$. We neglect the small mass difference between $J/\psi$
and $\eta_c$:
\begin{equation}
 m_h\approx m_{J/\psi} \approx m_{\eta_c}.
\end{equation}
From Eq.(4) it is easy to see that the helicity in the process is not conserved. For
helicity-conserving processes one can use the power-counting rule in \cite{FB}
to determine the asymptotic behavior of relevant form factors when $q^2\to\infty$.
For a process in which the helicity conservation is violated, one can use
the generalized power-counting rule in \cite{JMY} to determine
the asymptotic behavior of relevant form factors.
In our case one can obtain that ${\mathcal F}(s) \sim s^{-2}$ when $s\to\infty$.
\par

\begin{figure}[hbt]
\centering
\includegraphics[width=4cm]{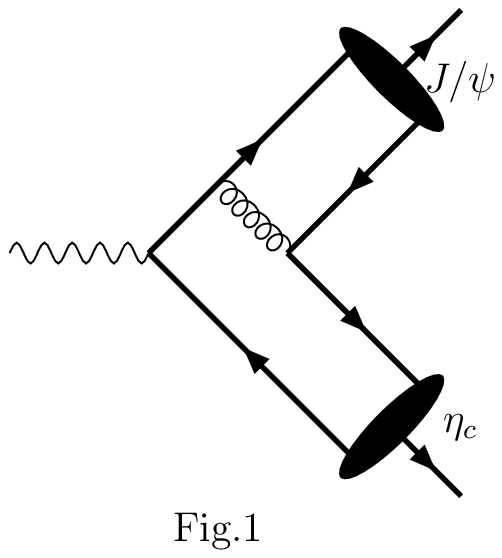}
\caption{One of the 4 Feynman diagrams  for the amplitude.}
\label{Feynman-dg1}
\end{figure}
\par
\par
At the leading order of $\alpha_s$ the contribution to the form factor
comes from four Feynman diagrams, one of them is given in Fig.1.
The contribution can be written as
\begin{eqnarray}
\langle J/\psi (p) \eta_c(k) \vert J^\mu \vert 0\rangle &=& \int
\frac{d^4 k_1}{(2\pi)^4} \frac{d^4 k_2}{(2\pi)^4}
   H^\mu_{ij,kl} (k_1,k_2,m_c) \int d^4 x e^{-ik_1\cdot x} \langle J/\psi(p)\vert \bar c_i (x) c_j(0) \vert 0\rangle
   \nonumber\\
   && \cdot \int d^4 y e^{-ik_2\cdot y} \langle \eta_c(k) \vert \bar c_k (y) c_l(0) \vert 0\rangle,
\end{eqnarray}
where the hard part $H^\mu_{ij,kl} (k_1,k_2,m_c)$ is
the amplitude for
\begin{equation}
\gamma^* (q) \to c_i(k_1) +c_k(k_2) + \bar c_j (p-k_1) +\bar
c_l(k-k_2)
\end{equation}
and all quarks can be off-shell. The hard part reads:
\begin{eqnarray}
H^\mu_{ij,kl} &=& \frac{i g_s^2}{(p-k_1+k_2)^2} \left ( \gamma_\nu
T^a \right )_{kj} \left ( \gamma^\nu T^a \frac{\gamma\cdot
(p+k_2)+m_c}{(p+k_2)^2-m_c^2}\gamma^\mu +\gamma^\mu
\frac{\gamma\cdot(k_1-q)+m_c}{(k_1-q)^2-m_c^2} \gamma^\nu T^a \right )_{il}
\nonumber\\
&& +\frac{i g_s^2}{(k-k_2+k_1)^2} \left ( \gamma_\nu T^a \right
)_{il} \left ( \gamma^\mu \frac{\gamma\cdot(k_2-q)+m_c}{(k_2-q)^2-m_c^2}
\gamma^\nu T^a
    +\gamma^\nu T^a \frac{\gamma\cdot (k+k_1)+m_c}{(k+k_1)^2-m_c^2}\gamma^\mu \right )_{kj}
\nonumber\\
\end{eqnarray}
where $ijkl$ are indices for color and spin.
\par
We will use light-cone wave-functions to describe the two
charmonia, in which a collinear expansion is used.
We take an coordinate system in which $J/\psi$ moves in
the $z$-direction and $\eta_c$ moves in $-z$-direction.
The momenta of these charmonia in the light-cone coordinate system
read:
\begin{eqnarray}
 p^\mu & =& (p^+, \frac{m^2_h}{2p^+},0,0)
 = p^+ l^\mu + \frac{ m_h^2 }{2p^+} n^\mu,
\nonumber\\
k^\mu & =& (\frac{m_h^2 }{2k^-},k^-, 0,0) =
\frac{m_h^2 }{2k^-} l^\mu + k^- n^\mu,
\end{eqnarray}
where $l^\mu =(1,0,0,0)$ and $n^\mu=(0,1,0,0)$ are two light-like
vectors. The transverse directions to the light-like directions
are denoted with the subscriber $\perp$.
In Eq.(7) the quark pair of $c(k_1) \bar c(p-k_1)$ is transited
into the $J/\psi$.
In general one expects that the dominant contributions
for the integration over $k_1$ are with $k_{1\perp}\sim \Lambda$ and
$k_1^- \sim \Lambda^2/2k_1^+$, where $\Lambda$ is the soft scale.
Similarly the dominant contributions for the integration over
$k_2$ are with $k_{2\perp}\sim \Lambda$ and
$k_2^+ \sim \Lambda^2/2k_2^-$.
Hence for these integrations one can expand the hard part
around $k_1^\mu =(k_1^+,0,0,0)$ and $k_2^\mu=(0,k_2^-,0,0)$
and in any soft scale:
\begin{eqnarray}
H^\mu_{ij,kl} (k_1,k_2, m_c) &\approx &  H^\mu_{ij,kl} (z_1 p^+ l, z_2k^-n,0)
+k^\rho_{1\perp}\left ( \frac{\partial H^\mu_{ij,kl} }{\partial k^\rho_{1\perp}}\right )
 (z_1 p^+ l, z_2 k^-n, 0)
\nonumber\\
&& +k^\rho_{2\perp}\left ( \frac{\partial H^\mu_{ij,kl} }{\partial k^\rho_{2\perp}}\right )
 (z_1 p^+ l, z_2 k^-n, 0)
 + m_c \left ( \frac{\partial H^\mu_{ij,kl} }{\partial m_c}\right )
 (z_1 p^+ l, z_2 k^-n, 0) +\cdots.
\end{eqnarray}
This is equivalent to expand the two matrix elements in Eq.(7) along light cones,
i.e., the matrix element $\langle \eta_c(k) \vert \bar c_k (y) c_l(0) \vert 0\rangle$
is expanded around $x^\mu =(0,x^-,0,0)$ and
$\langle \eta_c(k) \vert \bar c_k (y) c_l(0) \vert 0\rangle$
is expanded around $y^\mu =(y^+,0,0,0)$. However,
the expansion is not systematic in the sense
that the leading term in Eq.(11) will also lead to some contributions
which are at the same order of $\Lambda$ as those from higher orders.
The reason for this is clear: A Dirac field like $c(x)$ can be decomposed
along a light cone into a "good"- and "bad" component. The "bad" component
can be solved with the "good" component with equation of motion and
will lead to a contribution which is suppressed by $\Lambda$ in comparison
with that from the "good" component. This problem can be solved
by expanding the matrix elements
according to twists of operators. Matrix elements of operators
with a given twist are Fourier-transformed light-cone wave-functions.
For those matrix elements with light hadrons, the expansion in terms
of light-cone wave-functions have been studied in detail\cite{wfpion,wfrho}.
One  can use the results in \cite{wfpion,wfrho} to write down the expansion
for quarkonia. Up to twist-3 the expansion is\cite{cheng,wfpion,wfrho}:
\begin{eqnarray}
\langle \eta_c(k) \vert \bar c_k (y) c_l(0) \vert 0\rangle &=&
\frac{i}{12}f_{\eta_c} \Big \{  [  \gamma\cdot n \gamma_5 ]_{lk} k^-
                      \int d z_2 e^{i z_2 k^- y^+} \phi^{[2]} (z_2)
  -\frac{ m^2_{\eta_c}}{2m_c}  [  \gamma_5 ]_{lk}
    \int d z_2 e^{i z_2 k^- y^+} \phi^{[3]}_p (z_2)
\nonumber\\
 &&    + [ \sigma^{\mu\nu}\gamma_5 ]_{lk}
     n_\mu y_{\perp\nu} \int dz_2 e^{i z_2 k^- y^+} \phi^{[3]}_\sigma (z_2)\Big\}
     +\cdots,
\nonumber\\
\langle J/\psi(p)\vert \bar c_i (x) c_j(0) \vert 0\rangle &=&
\frac{1}{12} \Big \{
 i f_{J/\psi}^T  [ \sigma^{\mu\nu} ]_{ji}
 l_\mu \varepsilon^*_{\perp \nu}(p) p^+  \int dz_1 e^{i z_1 p^+ x^- }
 \psi_{\perp}^{[2]} (z_1)
\nonumber\\
&& +  f_{J/\psi} m_{J/\psi}   [ \gamma^\mu ]_{ji}
 \varepsilon^*_{\perp \mu}(p)   \int dz_1 e^{i z_1 p^+ x^- }
 \psi_{\perp}^{[3]} (z_1)\Big\}   +\cdots,
\end{eqnarray}
where the subscriber $\perp$ denote the transverse direction to the light-like
directions. It should be noted that
the space-time coordinate $x$ and $y$ in the matrix elements
is not on light-cones, in the side of  right-hand of the above equations
the expansion along light-cones is done.
The decay constants are defined as:
\begin{eqnarray}
\langle \eta_c(k) \vert \bar c (0)\gamma^\mu \gamma_5 c(0) \vert
0\rangle
   &=&  -i f_{\eta_c} k^\mu,
\nonumber\\
\langle J/\psi(p)\vert \bar c (0)\gamma^\mu c(0) \vert 0\rangle &=&
 f_{J/\psi} m_{J/\psi} \varepsilon^{*\mu}(p),\ \ \ \ \
 2f_{J/\psi} m_c = f_{J/\psi}^T m_{J/\psi}.
\end{eqnarray}
In Eq.(12) the numbers in the bracket $[\cdots]$ as
subscribers indicate twists.
The $\cdots$ denote twist-4 terms and those twist-3 terms
which are proportional to the factor
\begin{equation}
f_{J/\psi} - \frac{2m_c}{m_{J/\psi}} f_{J/\psi}^T = f_{J/\psi}
\left ( 1-\frac{4m_c^2}{m^2_{J/\psi}}\right ) \approx 0,
\end{equation}
which vanishes when the quark mass $m_c$ goes to infinite. We will
neglect contributions proportional to this factor. The above wave-functions
are normalized, i.e.,
\begin{equation}
\int_0^1 d z \{ \psi_\perp^{[2]},\psi_\perp^{[3]}, \phi^{[2]},
\phi_p^{[3]}, \phi_\sigma^{[3]} \} (z)  =1.
\end{equation}
\par
With the expansion in Eq.(12) one can calculate the form factor
in terms of these light-cone wave-functions.
If one only takes twist-2 wave-functions and
neglects the quark mass $m_c$, the form factor is zero, reflecting the fact
that the helicity is not conserved.  This also
implies that the contribution with twist-2 wave-functions only
is proportional to $m_c$
and it is at the same order of those contributions in which
one of twits-3 wave-functions is involved. We keep the contribution
of twist-2 by taking the finite quark mass into account.
It is straightforward to evaluate the form factor in terms
of these wave-functions. We obtain:
\begin{eqnarray}
{\mathcal F}(s) &=& \frac{8\pi \alpha_s (s) }{9} f_{\eta_c} f_{J/\psi}
\frac{1}{s^2}
         \int^1_0 d z_1 d z_2 \Big \{ \frac{- 2m_c^2}{m_{J/\psi} }
         \psi^{[2]}_\perp (z_1) \phi^{[2]}(z_2)
         \left [ \frac{1}{(1-z_1)^2 z_2} +\frac{1}{(1-z_2)z_1^2} \right ]
\nonumber\\
  && +m_{J/\psi} \psi_\perp^{[3]}(z_1) \phi^{[2]}(z_2)
  \left [ \frac{1}{z_2^2 (1-z_1)} -\frac{1}{z_2(1-z_1)} +\frac{1}{z_1(1-z_2)^2}
            -\frac{1}{z_1(1-z_2)} \right ]
 \nonumber\\
 && + \frac {2m^2_{\eta_c}}{m_{J/\psi}}\psi_\perp^{[2]}(z_1) \phi_p^{[3]} (z_2)
     \left [ \frac{1}{z_2(1-z_1)^2} +\frac{1}{z_1^2(1-z_2)} \right ] \Big\}
  \cdot \left (1+{\mathcal O} (\frac{\Lambda}{\sqrt{s}}) \right ).
\end{eqnarray}
It is interesting to note that the wave-function
$\phi_\sigma^{[3]}$ does not contribute at the considered order, as shown
by our calculation.
The correction to our result in Eq.(16) is suppressed
by the power of $\Lambda/\sqrt{s}$ or $\Lambda^2/s $, where $\Lambda$ can be the QCD parameter
$\Lambda_{QCD}$, the quark mass $m_c$ and masses of quarkonia.
\par
If we know these wave-functions we can give an numerical result for the form
factor and hence the cross-section. Unfortunately, these wave-functions are not
well known at the energy scale we are interested in. If the energy scale is very
large, these wave-functions approach to their asymptotic form:
\begin{eqnarray}
\phi^{[2]}(z) &\approx &\phi^{[3]}_\sigma(z) \approx  \psi_\perp^{[2]} (z) \approx 6
z(1-z),
\nonumber\\
\phi^{[3]}_p (z) & \approx & \psi^{[3]}_\perp (z) \approx 1.
\end{eqnarray}
If we take these asymptotic forms of wave-functions to make predictions, we
will have end-point singularities. These singularities may be regularized
by introducing a momentum cut. We regularize the end-point singularities
by change the integration range as:
\begin{equation}
\int_0^1 d z_1 \int_0^1 dz_2 \to \int_\varepsilon^{1-\varepsilon} d z_1
\int_\varepsilon^{1-\varepsilon} d z_2.
\end{equation}
with $\varepsilon = {m_c}/{\sqrt{s}}$.
For numerical predictions in this letter we take numerical values of parameters as:
\begin{eqnarray}
\sqrt{s}&=&10.6GeV, ~~~~~
\alpha_s(\sqrt{s})=0.1758,~~~
m_h\simeq 3.0GeV,~~m_c\simeq 1.6 GeV,\nonumber\\
f_{\eta_c} & \simeq & 350 MeV, ~~~~
f_{J/\psi}^T=\frac{2m_c}{m_h} f_{J/\psi},~~f_{J/\psi}\simeq 405MeV.
\end{eqnarray}
Taking the asymptotic form of the light-cone wave-functions and these parameters we obtain:
\begin{equation}
\sigma (e^+e^-\rightarrow J/\psi\,\eta_c)\simeq 1.31 {\rm fb}.
\end{equation}
\par
It is interesting to note that light-cone wave-functions as defined in Eq.(12) can be
calculated with NRQCD factorization, in which nonperturbative effects can be
parameterized with NRQCD matrix elements\cite{MaF}. It is easy to obtain the
leading order results as:
\begin{eqnarray}
\phi^{[2]}(z) = \phi^{[3]}_\sigma(z) = \psi_\perp^{[2]} (z) =
\phi^{[3]}_p (z) = \psi^{[3]}_\perp (z) =\delta(z -\frac{1}{2} ).
\end{eqnarray}
Using this type of wave-functions we obtain for the cross section:
\begin{equation}
\sigma(e^+e^-\rightarrow J/\psi\,\eta_c) \simeq  0.706 {\rm  fb}.
\end{equation}
All predictions in the above two cases are too small in comparison with the experimental result
in Eq.(1). However, the choice of forms of wave-functions in the two cases is not reasonable,
because the asymptotic form is only valid when the energy scale goes to infinity and the NRQCD predictions
in Eq.(20) are only reliable at the energy scale to be $m_c$ with possibly large
corrections from higher orders in $\alpha_s$ and relativistic corrections.
Here, we have an energy scale as $\sqrt{s}\approx 10$GeV,  which
is not close to $m_c$ and far from being infinity. In general, predictions are sensible to the
form of wave-functions.
Light-cone wave-functions are nonperturbative objects, which can be only determined with nonperturbative
methods or extracted from experimental results. The most extensively studied one is the wave-function
of $\pi$ and $\rho$(e.g., see \cite{wfpion,wfrho,Mpi}). Motivated by these studies,  we can make some models of
wave-functions for charmonia.
\par
A model for the twist-2
light-cone wave-function $\phi_\pi$ of $\pi$ was proposed long time ago in \cite{Mpi}, it
takes the form as
\begin{equation}
\phi_\pi (z) =6z (1-z) \Big\{1+\frac{3}{2} c \Big[5 (1-2z)^2-1\Big]\Big\},
\end{equation}
with $c=2/3$. A study with QCD sum rule gives $c=0.44$ at $\mu =1$GeV\cite{wfpion}. It should be noted that the shape
of $\phi_\pi$ with these nonzero values of $c$ is dramatically different than the shape with $c=0$, i.e., the shape
of asymptotic form. Motivated by this observation we assume
the twist-2 wave-function for $\eta_c$ to be
\begin{equation}
 \phi^{[2]}(z)=6z (1-z) \Big\{1+0.44 \frac{3}{2} \Big[5 (1-2z)^2-1\Big]\Big\}.
\end{equation}
The twist-3 wave-functions of $\pi$ are also studied in \cite{wfpion}. Through a study of
recursion relations of moments of $\phi_p^{[3,\pi]} (z) $ and with QCD sum rule the form
of $\phi_p^{\pi} (z)$ is determined at $\mu =1$GeV as:
\begin{equation}
\phi_p^{[3,\pi]}=1+(0.39-2.5\rho_{\pi}^2) C_2^{1/2}(2z-1)+
                 (0.117-4.914 \rho_{\pi}^2) C_4^{1/2}(2z-1),
\end{equation}
with $\rho_\pi =(m_u+m_d)^2/m_\pi^2$. $C_n^{\lambda} (x)$ denotes Gegenbauer polynomials.
It should be noted
that the terms with $\rho_\pi$ represent the part of meson-mass correction. This part
can be totally different in the case of $\eta_c$. Based on this fact
we assume the wave-function $\phi_p^{[3]}$
to be the form:
\begin{equation}
\phi_p^{[3]} (z) = 1 +(0.39 -2.5 \rho^2_{\eta_c} ) C_2^{1/2}(2z-1)
                   +(0.117-4.914\rho^2_{\eta_c} ) C_4 ^{1/2} (2z-1),
\end{equation}
where we simply replace $\rho_{\pi}$ with $\rho_{\eta_c}$ and $\rho_{\eta_c}$
takes the form as $4 m_c^2 a/M^2_{\eta_c}$ with
a free parameter $a$.
With similar technics the wave-functions for $\rho$ is determined at $\mu=1$GeV
as\cite{wfrho}:
\begin{eqnarray}
\psi_\perp^{[2,\rho]} (z) &=& 6 z(1-z) ( 1+ 0.3(5(2z-1)^2-1)),
\nonumber\\
\psi_{\perp}^{[3,\rho]}(z) &=& 1-1.6248 C_2^{1/2}(2z-1)-0.413 C_4^{1/2}(2z-1),
\end{eqnarray}
We assume the wave-functions of $J/\psi$ to be:
\begin{equation}
\psi^{[2]} (z) =\psi_\perp^{[2,\rho]} (z), \ \ \\ \ \
\psi_\perp^{[3]} (z) =\psi_\perp^{[3,\rho]} (z).
\end{equation}
With these wave-functions in Eq.(23,25,27) we obtain the cross-section for differen values
of $a$:
\begin{eqnarray}
\sigma(e^+e^-\rightarrow J/\psi\,\eta_c) & \simeq &  7.37 {\rm  fb}, \  \
{\rm for\ } a=1,
\nonumber\\
\sigma(e^+e^-\rightarrow J/\psi\,\eta_c) & \simeq &  20.1 {\rm  fb}, \  \
{\rm for\ } a=1.5,
\nonumber\\
\sigma(e^+e^-\rightarrow J/\psi\,\eta_c) & \simeq &  31.7 {\rm  fb}, \  \
{\rm for\ } a=1.75.
\end{eqnarray}
The cross-section increases with increasing $a$. The predicted cross section
with $a=1.75$ is much larger than that predicted with NRQCD approach and is
more comparable with experimental result in Eq.(1). It should be emphasized that
the forms of used wave-functions are assumed without
any solid arguments, although it is motivated with those of $\pi$ and $\rho$.
In this model we neglect the evolution effects of light-cone wave-functions. 
The numbers in the light-cone wave-functions we use are calculated
with the QCD sum rules for $\pi$ and $\rho$ at $\mu=1$GeV. The physics
reflected by these numbers are actually nothing to do the physics of charmonia.
However, it shows the possibility to obtain a large cross-section at the order
of the experimentally observed with our approach.
A detailed study of these light-cone wave-functions of charmonia
is needed to obtain a reliable prediction.
\par
To summarize: We have studied the exclusive production of $e^+e^-\to J/\psi \eta_c$,
in which we have taken charm quarks as light quarks and used light-cone wave-functions
to parameterize nonperturbative effects related to charmonia.
In comparison with NRQCD factorization,
the factorization of our approach may be achieved in a more clean way and the perturbative
coefficients will not have corrections with large logarithms like $\ln(\sqrt{s}/m_c)$
from higher orders, while in the approach of NRQCD factorization, these large logarithms
exist and call for resummation. The forms of these light-cone wave-functions
are known if the energy scale is close to $m_c$ or is very large.
Unfortunately, these wave-functions at the considered energy scale,  which is
not close to $m_c$ and far from being very large, are unknown.
With a simple model
of light-cone wave-functions, we are able to predict the cross-section which
is at the same order of that measured by Belle.
But, this model may not represent the physics of charmonia. A systematic study
of these light-cone wave-functions is required to have a precise prediction.
\par\vskip20pt
Acknowledgment:
\par
This work is  supported by National
Natural Science Foundation of P.R. China.

\par\vfil\eject


\end{document}